\documentclass[preprint,review,3p,12pt]{elsarticle}

\usepackage{amssymb}
\journal{Journal of Crystal Growth}

\begin{document}

\begin{frontmatter}



\title{Growth of (CH$_3$)$_2$NH$_2$CuCl$_3$ single crystals using evaporation method with different temperatures and solvents}


\author[1,2]{L. M. Chen}
\author[2]{W. Tao}
\author[2]{Z. Y. Zhao}
\author[2]{Q. J. Li}
\author[2]{W. P. Ke}
\author[2]{X. M. Wang}
\author[2]{X. G. Liu}
\author[2]{C. Fan}
\author[2]{X. F. Sun\corref{cor1}}

%\footnote{Correspondence author. Tel.: 86-551-3600499, Fax:
%86-551-3600499.\\ Postal address: No.96 Jinzhai Road, Hefei, Anhui
%230026, China.\\ Email address: xfsun@ustc.edu.cn}}

\address[1]{Department of Physics, University of Science and
Technology of China, Hefei, Anhui 230026, China}

\address[2]{Hefei National Laboratory for Physical Sciences at Microscale,
University of Science and Technology of China, Hefei, Anhui
230026, China}

\cortext[cor1]{Corresponding author. Tel.: 86-551-3600499, Fax:
86-551-3600499.\\ Email address: xfsun@ustc.edu.cn}

%\fntext[label*]{Postal address: No.96 Jinzhai Road, Hefei, Anhui
%230026, China.}

\date{\today}

\begin{abstract}

The bulk single crystals of of low-dimensional magnet
(CH$_3$)$_2$NH$_2$CuCl$_3$ (DMACuCl$_3$ or MCCL) are grown by a
slow evaporation method with different kinds of solvents,
different degrees of super-saturation of solution and different
temperatures of solution, respectively. Among three kinds of
solvent, methanol, alcohol and water, alcohol is found to be the
best one for growing MCCL crystals because of its structural
similarity to the raw materials and suitable evaporation rate. The
best growth temperature is in the vicinity of 35 $^{\circ}$C. The
problem of the crystals deliquescing in air has been solved
through recrystallization process. The crystals are characterized
by means of x-ray diffraction, specific heat and magnetic
susceptibility.

\end{abstract}


\end{frontmatter}

% \linenumbers


\newpage

\section{Introduction}

Catena (dimethylammonium-bis($\mu_2$-chloro)-chlorocuprate)
(CH$_3$)$_2$NH$_2$CuCl$_3$ (DMACuCl$_3$ or MCCL) is an
organic-metallic magnetic material with quasi-one-dimensional
alternating antiferromagnetic-ferromagnetic (AFM-FM) Heisenberg
chain ($S$ = 1/2). MCCL is very interesting in the regard that it
has a rather complex $H$-$T$ phase diagram and multiple
magnetic-field-induced phase transitions. Willett first proposed
that Cu$^{2+}$ ions ($S$ = 1/2) coupled via Cu-halide-Cu bridges
to form magnetic chains along the crystallographic $a$ axis, but
Stone {\it et al.} recently found that there was a significant
dispersion along the $b$ axis by inelastic neutron scattering,
indicating that the $b$ axis is the one-dimensional magnetic axis
\cite{Willett, Willett2, Stone, Abouie}. It is found that the
magnetization shows a 1/2 plateau, corresponding to a
field-induced gapped state, with magnetic field range from 2 to
3.5 T \cite{Inagaki}. There are a spontaneous AFM ordering and a
field-induced one below and above the plateau, respectively
\cite{Inagaki2, Yoshida}. It would be very useful for studying the
mechanisms of these magnetic transitions if one can get
high-quality single crystals of MCCL.

A slow evaporation method using water, methanol or alcohol as a
solvent was originally introduced in 1960's to grow MCCL crystals
\cite{Willett}. Recently, several groups also obtained MCCL
crystals, including the deuterated crystals, by using Willett's
method \cite{Willett2, Stone, Inagaki, Inagaki2, Yoshida}.
However, the conditions of the crystal growth, the sizes and the
quality of the obtained crystals are not clearly mentioned in
these papers. It is therefore still important to investigate the
appropriate growth procedure. In this work, we report a detailed
study on the MCCL crystal growth using this method. It is found
that the MCCL single crystals can be synthesized in different
kinds of solvents, including methanol, alcohol and water. The
sizes and the shapes of the crystals, however, are dependent not
only on the type of solvent but also on the degree of
super-saturation of solution and the growing temperature. The
biggest crystal with size of 12$\times$7$\times$4 mm$^3$ is
obtained from 100 ml alcohol solution (containing raw materials of
15 mmol CuCl$_2$$\cdot$2H$_2$O and 15 mmol (CH$_3$)$_2$NH$_2$Cl)
at $35\,^{\circ}\mathrm{C}$. A common problem of the crystal
deliquescing in air can be overcome by using a recrystallization
process.

\section{Crystal Growth}
(CH$_3$)$_2$NH$_2$Cl and CuCl$_2$$\cdot$2H$_2$O in the molar ratio
1:1 are used as the starting materials for synthesizing MCCL. The
chemical reaction formula between the two raw materials is

(CH$_3$)$_2$NH$_2$Cl+CuCl$_2\cdot$2H$_2$O$\rightarrow$(CH$_3$)$_2$NH$_2$CuCl$_3$+H$_2$O.

At first, 15 mmol CuCl$_2\cdot$2H$_2$O is dissolved in a selected
solvent (methanol, water or alcohol) at different temperatures.
Then, the same mole amount of (CH$_3$)$_2$NH$_2$Cl is added to the
solution slowly with continuously stirring. Keeping the solution
in the super-saturation state, then the crystals can grow
gradually with the solution continuously evaporating.

We study the influence of the temperature on the crystal growth
under the condition that both the amount of raw material and the
degree of super-saturation solution are invariant. The volumes of
solvents are chosen to be 150 ml. For aqueous solvent, the
solution evaporates too fast to form single crystals at
temperatures above 50 $^{\circ}$C, while we can obtain shining
tiny slice-like crystals and big crystals with thin smooth surface
at 35 $^{\circ}$C and 20 $^{\circ}$C, respectively. In addition,
the solution at 20 $^{\circ}$C evaporates very slowly and it takes
a very long time (about three months) to get crystals. These
results are listed in Table 1. The similar investigations and
results by using alcohol as the solvent are given in Table 2.
There are no crystals at temperatures above 40 $^{\circ}$C; while
many high-quality single crystals with typical size of
(3--7)$\times$(2--3)$\times$(2--4) mm$^3$ can be obtained after
one week at 35 $\pm$ 2 $^{\circ}$C. If we keep the growing
temperature at 20 $^{\circ}$C, the obtained crystals have
considerably big sizes but poor quality, checked by the x-ray
diffraction (XRD). For methanol solvent, there are no any
high-quality single crystals though we make many attempts, see
results in Table 3. From the above, there is notable difference
among the results of three solvents at the same growth
temperatures. The main reason is that the ability of crystal
growth may strongly depend on the similarity of structure between
the solute and the solvent. Another possible reason is that the
different solvent molecules have different adsorption choices to
the faces of the crystals, leading to different growth rates of
the crystallographic planes and consequently forming different
shapes of the crystals \cite{Pamplin, Christensen, Brice}.
Moreover, the growth temperatures have significant effect on the
quality and shape of crystals even in the same solvent. Figure 1
shows the shapes of MCCL crystals grown at different temperatures
in alcohol. The temperature can be considered as an activation
energy to affect the processes of the crystal growth, which is
either a pure surface reaction process or a pure diffusion
process. In general, the surface reaction of the crystallization
process plays the main role at low temperature. When the
temperature rises, the growth rate becomes faster and the
diffusion process will gradually dominate the growth of the
crystals. The crystals grown at an appropriate higher temperature
usually has better quality than those grown at low temperature
(but too high temperature causes too fast evaporation to generate
crystals), because the driving force of the crystallization, which
is the ability of crystallized particles rejecting impurities, is
enhanced with increasing temperature \cite{Singh}. Generally, the
sample has a flat irregular shape and normally grows with the
(110), (011) and (001) planes developed. The largest crystal face
is the (110) plane.

One problem we meet is that the MCCL single crystals deliquesce in
air so easily that they can completely be destroyed after keeping
in air for two days, which makes both the sample preservation and
the measurement very difficult. We adopt the following method to
solve this problem. After the first-run growth is nearly finished,
a same amount of solvent is added to the beaker, which contains
the crystals. The solution is kept at the same condition and then
the crystals are partially dissolved and re-crystallized. The
re-crystallized single crystals have the similar shapes to those
from the first run but do not deliquesce any more. It is likely
that the purity of the synthesized crystals can be improved by the
recrystallization process.

The different degree of super-saturation of the solution also
affects the crystal quality. As shown in Fig. 2, the ratio of 15
mmol raw materials to 80 ml alcohol makes the solution at an
unstable super-saturated state and can not form good crystals.
From the solution of 15 mmol raw materials dissolved into 100 ml
alcohol, the obtained crystals are very large and the biggest one
reaches 230 mg. However, the crystal size becomes smaller with
further reducing the degree of super-saturation, as shown in Fig.
2(c). This result indicates that the initial concentration of the
solution is very important in determining the crystals formation
and their growth.

To sum up, we compare the results of the crystal growth from
different types of solvent, different temperatures of solution and
different degrees of super-saturation solution, respectively. We
obtain the best crystal growth condition: prepare a
super-saturated solution with the ratio of 1 mmol raw materials to
10 ml solvent and grow crystals by continuously evaporating the
solution at 35 $^{\circ}$C. The alcohol is a much better solvent
than water and methanol.

\section{Structural characterization}

%\subsection{\bf Orbital ordering}

Figure 3(a) shows the x-ray powder diffraction pattern for the
crystals we obtained from the best growing procedure discussed
above. The data are consistent with the monoclinic structure
(space group I2/$a$) with lattice parameters $a$ = 11.97 \r{A},
$b$ = 8.6258 \r{A}, $c$ = 14.34 \r{A}, and $\beta$=
97.47$^{\circ}$ \cite{Stone}. Figure 3(b) shows a rocking curve of
(002) diffraction for a selected single crystals. The peak is very
narrow with the full width at half maximum of 0.07$^{\circ}$,
indicating a high quality of this sample.

\section{Magnetic susceptibility and specific heat}

Figure 4 shows the temperature dependence of magnetic
susceptibility measured in 4 T magnetic field perpendicular to the
$ab$ plane using a SQUID-VSM (Quantum Design). The susceptibility
increases monotonically with decreasing temperature, showing a
Curie-like behavior \cite{Ajiro}. There is no magnetic transition
down to 2 K, which is consistent with the former studies.

Specific heat is measured by the Physical Property Measurement
System (PPMS, Quantum Design). Figures 5(a), 5(b) and 5(c) show
the effects of magnetic field on $C$($T$) for three characteristic
field ranges, which are separated by the 1/2 magnetization
plateau. In general, all these data are consistent with those in a
former report \cite{Yoshida}. The temperature dependence of the
specific heat at low field has a well-defined peak in the low
temperature regime, as shown in Fig. 5(a). In zero field, there is
a sharp peak at the N$\acute{e}$el transition temperature $T_N$ =
0.8 K. The height of the peak decreases gradually with increasing
field, indicating the suppression of AFM ordering by the external
field. The peak disappears completely in the 1/2-plateau field
range (2--3.5 T), as shown in Fig. 5(b), suggesting a transition
from the AFM magnetic ordering into the field-induced gapped
state. When the field is above 3.5 T, another peak appears and
increases rapidly with increasing field, as shown in Fig. 5(c),
indicating that the spin gap is closed and the system turns into
the field-induced magnetic ordering state. All such peculiarities
originate most likely from the interplay between the coexisting FM
and AFM dimers \cite{Abouie, Inagaki, Inagaki2, Yoshida}.

Therefore, both the magnetic susceptibility and the specific heat
measurements of our MCCL singe crystals are essentially same as
those reported in earlier papers \cite{Abouie, Inagaki, Inagaki2,
Yoshida}, in which these physical properties have already been
discussed in details. In this regard, the quality of our crystals
are also confirmed and it is possible to carry out more
experimental investigations, like low-temperature heat transport
measurements, by using these high-quality crystals.

\section{CONCLUSIONS}

MCCL single crystals are grown at different temperatures in
alcohol, aqueous and methanol solutions, respectively. The most
appropriate condition for MCCL crystal growth is starting from a
super-saturation solution with ratio of 1 mmol each raw material
to 10 ml alcohol solvent and keeping the growth temperature at 35
$^{\circ}$C. The problem of the crystals deliquescing is solved by
recrystallization method. The x-ray diffraction data indicate
single phase and good crystallinity of the obtained crystals. The
magnetization measurement confirms the absence of magnetic
ordering above 2 K, while the specific heat measurements at very
low temperatures and in high magnetic fields demonstrate the
low-field AF ordering, the field-induced spin gapped state and the
field-induced magnetic ordering upon increasing field.

\section*{ACKNOWLEDGMENTS}

This work was supported by the Chinese Academy of Sciences, the
National Natural Science Foundation of China and the National
Basic Research Program of China (Grant Nos. 2009CB929502 and
2006CB922005).

%\label{}



%% References


\bibliographystyle{elsarticle-num}
\bibliography{<your-bib-database>}

\begin{thebibliography}{}

\bibitem{Willett}
R. D. Willett, J. Chem. Phys. {\bf 44}, 39 (1966).

\bibitem{Willett2}
R. D. Willett, B. Twamley, W. Montfrooij, G. E. Granroth, S. E.
Nagler, D. W. Hall, J.-H. Park, B. C. Wastson, M. W. Meisel, and
D. R. Talham, Inorg. Chem. {\bf 45}, 7689 (2006).

\bibitem{Stone}
M. B. Stone, W. Tian, M. D. Lumsden, G. E. Granroth, D. Mandrus,
J.-H. Chung, N. Harrison, and S. E. Nagler, Phys. Rev. Lett. {\bf
99}, 087204 (2007).

\bibitem{Abouie}
J. Abouie and S. Mahdavifar, Phys. Rev. B {\bf 78}, 184437 (2008).

\bibitem{Inagaki}
Y. Inagaki, A. Kobayashi, T. Asano, T. Sakon, H. Kitagawa, M.
Motokawa, and Y. Ajiro, J. Phys. Soc. Jpn. {\bf 74}, 2683 (2005).

\bibitem{Inagaki2}
Y. Inagaki, O. Wada, K. Ienaga, H. Morodomi, T. Kawae, Y. Yoshida,
T. Asano, Y. Frukawa, and Y. Ajiro, J. Phys.: Conf. Ser. {\bf
150}, 042067 (2009).

\bibitem{Yoshida}
Y. Yoshida, O .Wada, Y. Inagaki, T. Asano, K. Takeo, T. Kawae, K.
Takeda, and Y. Ajiro, J. Phys. Soc. Jpn. {\bf 74}, 2917 (2005).

\bibitem{Pamplin}
B. R. Pamplin, {\it Crystal Growth} (Pergamon Press, Oxford, New
York Toronto, Sydney, 1975).

\bibitem{Christensen}
A. N. Christensen, F. Leccabue, C. Paorici, O. Vigil, {\it Crystal
Growth and Characterization of Advanced Materials} (World
Scientific, Singapore, New Jersey, 1987).

\bibitem{Brice}
J. C. Brice, {\it The Growth of Crystals from Liquid} (American
Elsevier Publishing Company, Inc. New York, 1973).

\bibitem{Singh}
N. B. Singh, T. Henningsen, R. H. Hopkins, R. Mazelsky, R. D.
Hamacher, E. P. Supertzi, F. K. Hopkins, D. E. Zelmon, O. P.
Singh, J. Cryst. Growth {\bf 128}, 976 (1993).

\bibitem{Ajiro}
Y. Ajiro, K. Takeo, Y. Inagaki, T. Asano, A. Shimogai, M. Mito, T.
Kawae, K. Takeda, T. Sakon, H. Nojiri, and M. Motokawa, Physica B
{\bf 329-333}, 1008 (2003).

\end{thebibliography}





\newpage
\begin{figure}
\includegraphics[width=10cm]{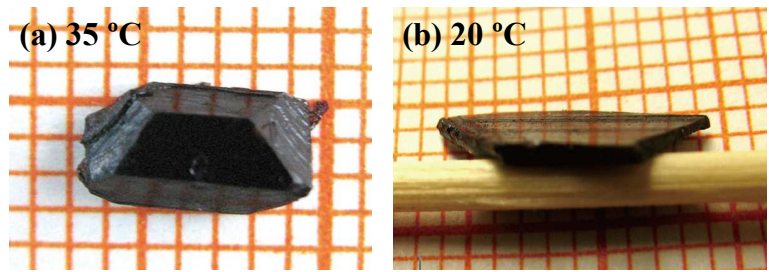}
 \caption{The crystals grown at different temperatures in alcohol.
They have very different shapes. In panel (a), the irregular
crystal has four bright planes and good quality. In panel (b), the
flat crystal has a big shining plane but poor quality.}
\end{figure}

 $ $
\newpage

\begin{figure}
\includegraphics[width=15cm]{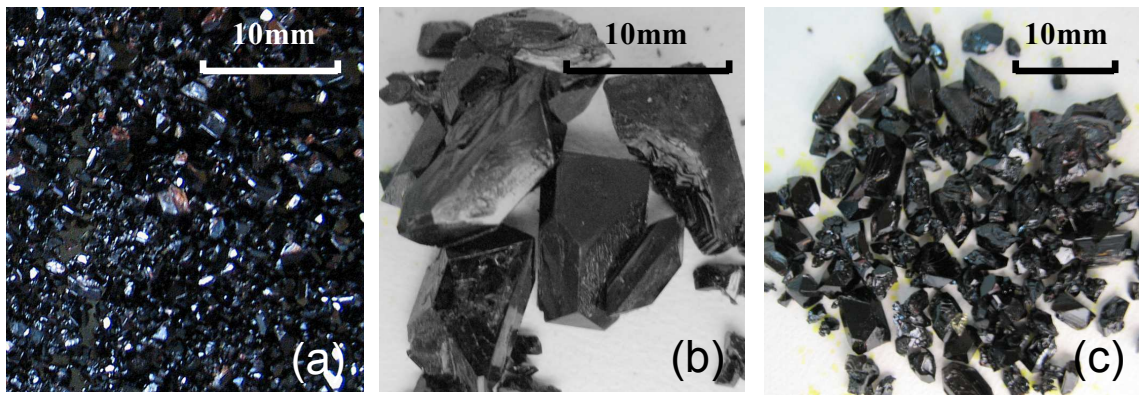}
 \caption{Crystals grown from alcohol solvent at
35 $^{\circ}$C with different amounts of solvent. 15 mmol of each
raw material is dissolved in 80 ml, 100 ml and 150 ml alcohol,
respectively.}
\end{figure}

$ $
\newpage

\begin{figure}
\includegraphics[width=15cm]{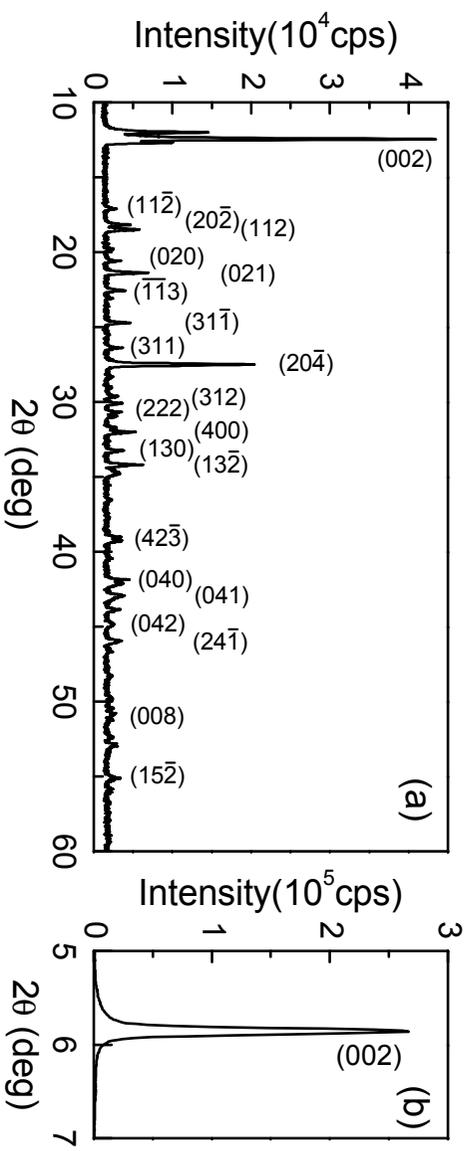}
 \caption{(a) X-ray powder diffraction pattern. (b) Rocking curve of
(002) peak for a piece of MCCL single crystal.}
\end{figure}

$ $
\newpage
\begin{figure}
\includegraphics[width=8.5cm]{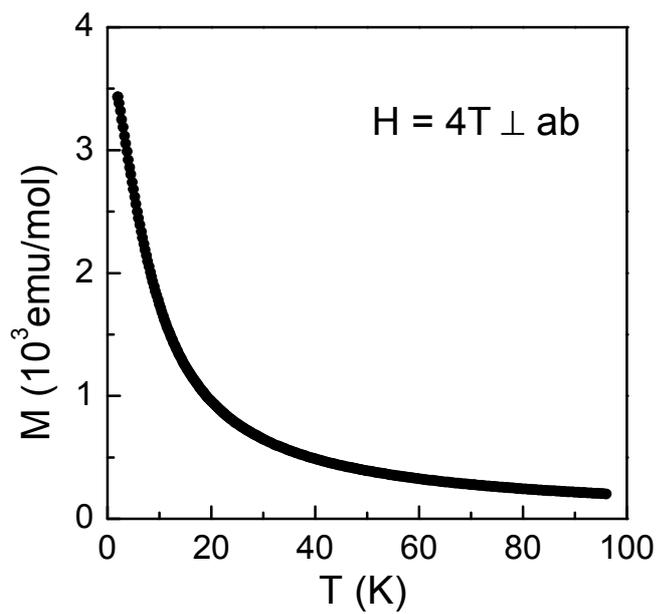}
 \caption{Temperature dependence of magnetic susceptibility measured
in 4 T magnetic field perpendicular to the $ab$ plane.}
\end{figure}

$ $
\newpage
\begin{figure}
\includegraphics[width=8.5cm]{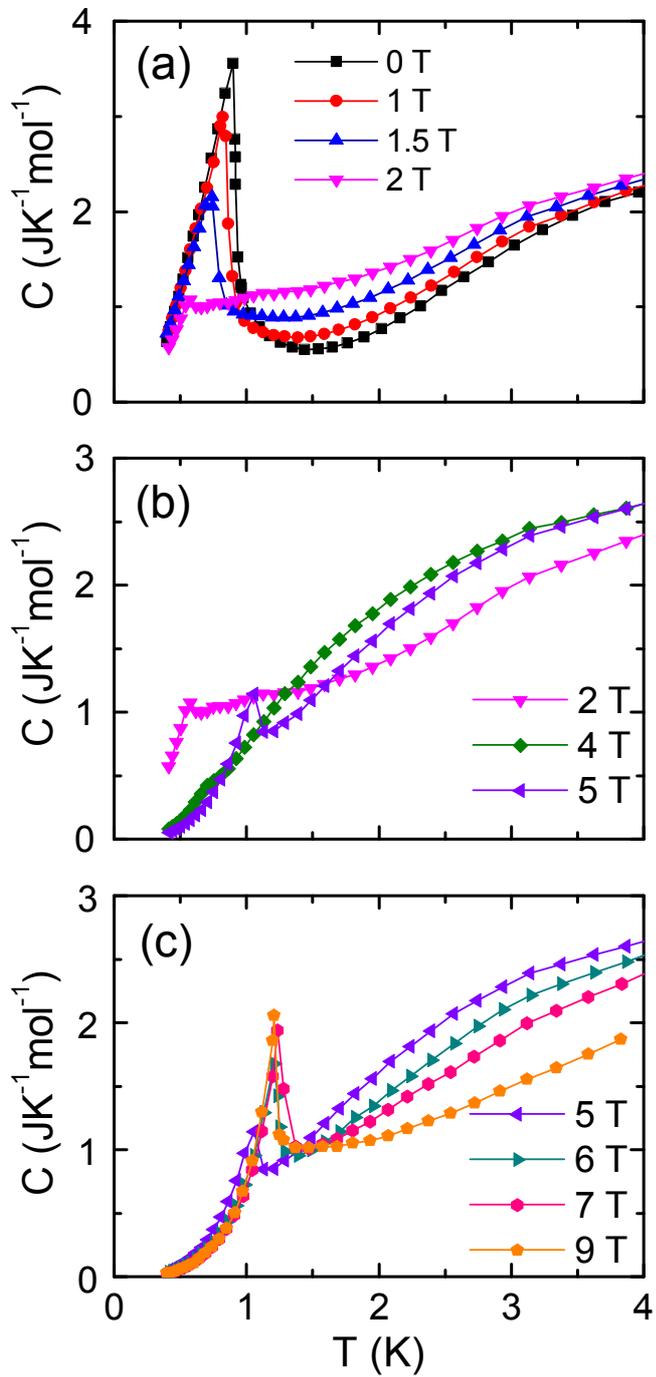}
 \caption{Specific heat of MCCL crystal as a function of temperature
for magnetic field applied perpendicular to the $ab$ plane.}
\end{figure}
\clearpage

\newpage

%---------------------------------------Table. 1----------------------------------------------------
\begin{table}[htbp]
\caption {Results for different growth temperatures in 150 ml
solvent of water.} \centering
\begin{tabular}{p{80pt}p{250pt}}
\hline \hline
Temperature & Crystal State\\
\hline
20 $^{\circ}$C  & big bright pieces, size about (5--9)$\times$(3--5) mm$^2$\\
35 $^{\circ}$C  & bright tiny slice \\
40 $^{\circ}$C  & none \\
50 $^{\circ}$C  & none \\
\hline \hline
\end{tabular}
\end{table}
\clearpage

\newpage

%---------------------------------------Table. 2----------------------------------------------------
\begin{table}[htbp]
\caption {Results for different growth temperatures in 150 ml
solvent of alcohol.} \centering
\begin{tabular}{p{80pt}p{320pt}}
\hline \hline
Temperature & Crystal State\\
\hline
20 $^{\circ}$C  & flat crystal, but poor quality \\
35 $^{\circ}$C  & high quality crystals, size about (3--7)$\times$(2--3)$\times$(2--4) mm$^3$\\
40 $^{\circ}$C  & none \\
50 $^{\circ}$C  & none \\
\hline \hline
\end{tabular}
\end{table}
\clearpage

\newpage

%---------------------------------------Table. 3----------------------------------------------------
\begin{table}[htbp]
\caption {Results for different growth temperatures and different
volumes for methanol solvent. The amount of starting raw materials
is 15 mmol.} \centering
\begin{tabular}{p{80pt}p{80pt}p{280pt}}
\hline \hline
Volume & Temperature & Crystal State\\
\hline
60 ml & 20 $^{\circ}$C  & thin bright pieces, size about (4--7)$\times$(2--3) mm$^2$ \\
70 ml & 20 $^{\circ}$C  & small crystals, size is (0.5--2)$\times$(0.3--1)$\times$(0.1--0.8) mm$^3$\\
150 ml & 20 $^{\circ}$C  & bright thin pieces, size about 5$\times$4 mm$^2$  \\
150 ml & 35 $^{\circ}$C  & none \\
150 ml & 35 $^{\circ}$C  & none \\
150 ml & 40 $^{\circ}$C  & none \\
150 ml & 50 $^{\circ}$C  & none \\
\hline \hline
\end{tabular}
\end{table}
\clearpage

\end{document}